\documentclass[%
mathleft,%
final,%
]{an}
\usepackage{graphicx}
\usepackage{times}
\begin{document}

% The following seven commands are intended for editorial usage and
% should be ignored by the author(s).
\Pagespan{1}{}% Document's page range.
% If second parameter is left empty, the last page is computed
% automatically.
\Yearpublication{2013}%
\Yearsubmission{2012}%
\Month{1}%
\Volume{334}%
\Issue{1}%
\DOI{This.is/not.aDOI}%

\title{Comparison between accretion-related properties of Herbig Ae/Be and T Tauri stars}

\author{I. Mendigut\'\i{}a\thanks{\email{Ignacio.Mendigutia@cab.inta-csic.es}}}
% Example for footnote, note the usage of the \fnmsep command
% as separator between institute number and footnote mark}

\titlerunning{Comparison between accretion-related properties of Herbig Ae/Be and T Tauri stars}
\authorrunning{I. Mendigut\'\i{}a}
\institute{Departamento de F\'{\i}sica Te\'{o}rica, M\'{o}dulo 15,
     Facultad de Ciencias, Universidad Aut\'{o}noma de Madrid, PO Box
     28049, Cantoblanco, Madrid, Spain.\\
}

\received{XXXX}
\accepted{XXXX}
\publonline{XXXX}


\abstract{This paper summarizes several results concerning the comparison between accretion-related properties of cool (T Tauri; T $<$ 7000 $K$, M $<$ 1 M$_{\odot}$) and hot (Herbig Ae/Be; 7000 $<$ T($K$) $<$ 13000; 1 $<$ M(M$_{\odot}$) $<$ 6) pre-main sequence (PMS) stars. This comparison gives insight into the analogies/differences in the physics of the star-disk interaction and in the physical mechanisms driving disk dissipation.\\ 
Several optical and near-IR line luminosities used for low-mass objects are also valid to estimate typical accretion rates for intermediate-mass stars under similar empirical expressions. In contrast, the H$\alpha$ width at 10$\%$ of peak intensity is used as an accretion tracer for T Tauris, but is not reliable to estimate accretion rates for Herbig Ae/Bes. This can be explained as a consequence of the different stellar rotation rates that characterize both types of stars. In addition, there are similar trends when the accretion rate is related to the near-IR colours and disk masses, suggesting that viscous accretion disk models are able to explain these trends for both T Tauri and Herbig Ae/Be stars. However, there are two major differences between cool and hot PMS objects. First, the inner gas dissipation timescale, as estimated by relating the accretion rates and the stellar ages, is slightly faster for Herbig Ae/Be stars. This could have implications on the physical mechanism able to form planets around 
objects more massive than the Sun. Second, the relative position of Herbig Ae/Bes with disks showing signs of inner dust clearing in the accretion rate--disk mass plane contrasts with that of transitional T Tauri stars, when both samples are compared with ''classical'', non-evolved disks. This latter difference could be pointing to a different physical mechanism driving disk evolution, depending on the stellar regime.}

\maketitle

\section{Introduction}

The star formation process can be roughly divided into different stages, from the initial collapse of a dense molecular core, passing through an optically-visible pre-main sequence (PMS) phase, to the final MS star eventually surrounded by a planetary system and/or debris disk. The PMS phase plays a key role in our understanding of planet formation, given that planets are in principle formed in the circumstellar ``protoplanetary'' disks that surround young stars with masses that range from $\sim$ 0.1 to $\sim$ 8--10 M$_{\odot}$. However, most studies on PMS objects focused on the low-mass T Tauri (TT) stars, and much less is known about their massive counterparts, the Herbig Ae/Be (HAeBe) stars.\\ 

The mass accretion rate ($\dot{M}_{\rm acc}$) measures the amount of gas transferred from the inner disk to the stellar surface per time unit. This parameter constitutes an indirect estimator of the amount of gas in the inner disk, and can be used to provide an alternative estimate of the disk mass (M$_{\rm disk}$) (Hartmann et al. \cite{Hartmann98}). Moreover, the study of PMS stars in the $\dot{M}_{\rm acc}$--M$_{\rm disk}$ plane provides hints on the physical mechanisms that control the star-disk interaction and drive disk dissipation (Alexander \& Armitage \cite{Alexander07}). For these and other reasons, a large number of works deal with accretion in low-mass PMS stars (see e.g. the review in Calvet, Hartmann \& Strom \cite{Calvet00} and references therein). It was more recently when efforts are being focused also on the HAeBe regime.\\
   
This paper summarizes recently obtained results on several accretion-related properties of HAe and late-type HBe stars (Mendigut\'{\i}a et al. \cite{Mendigutia11a}, \cite{Mendigutia11b}; Paper I and Paper II hereafter), compared to those previously derived for lower-mass TT stars (Mendigut\'{\i}a et al. \cite{Mendigutia12}; Paper III hereafter). Section \ref{Sect: Estimating} summarizes the methodologies applied to estimate accretion rates. Sections \ref{Sect: Accretiontraz}, \ref{Sect: Accretionevol} and \ref{Sect: Accretioparam} include some results on the accretion tracers, as well as several accretion properties related to stellar and disk parameters. Section \ref{Sect: Conclusions} includes the main conclusions.

\section{Accretion rate estimates}
\label{Sect: Estimating}
Magnetospheric accretion (MA hereafter; Uchida \& Shibata \cite{Uchida85}, K\"onigl \cite{Konigl91}, Shu et al. \cite{Shu94}) is the accepted paradigm that explains disk to star accretion in TT stars. In this model, the gas in the inner disk is channeled through the stellar magnetic field lines until it reaches the stellar surface, generating hot accretion shocks. Accretion rate estimates in TT stars are based on MA, mainly by reproducing the spectroscopic veiling, line profiles and UV continuum excess that these objects show (Calvet et al. \cite{Calvet00}). Fortunately, that type of complex modelling is not always necessary since the accretion luminosity correlates with the luminosity of several emission lines that span from the UV to the near-IR (see e.g. Rigliaco et al. \cite{Rigliaco11} and references therein, and Sect. \ref{Sect: Accretiontraz}), which are used as empirical tracers of accretion.\\ 

Regarding the HAeBes, there is an open debate on the origin of the magnetic fields that several of these objects show, and on whether they are strong enough to drive MA (Wade et al. \cite{Wade07}, Hubrig et al. \cite{Hubrig09}). However, rising consensus on the applicability of MA is emerging, at least for several HAeBe stars, mainly HAe and late-type HBe objects (e.g. Vink et al. \cite{Vink02}, Muzerolle et al. \cite{Muzerolle04}, Mottram et al. \cite{Mottram07}, Donehew \& Brittain \cite{DonBrit11}, Paper II). Accretion rate estimates for HAeBe stars are scarce, compared with those for TTs. The first results for a wide sample of HAeBes were carried out in Garcia Lopez et al. (\cite{GarciaLopez06}), by using the calibration with the Br$\gamma$ emission line obtained by Calvet et al. (\cite{Calvet04}). This calibration was derived from intermediate-mass T Tauri stars with stellar temperatures lower than 6300 $K$, and was extrapolated with uncknown accuracy to hotter HAeBe stars by Garcia Lopez et al. (\cite{GarciaLopez06}). More recently, Donehew \& Brittain (\cite{DonBrit11}) measured the excess in the Balmer region of the spectra shown by a sample of 33 HAeBes. The Balmer excesses were associated to accretion rates from the calibration in Muzerolle et al. (\cite{Muzerolle04}), which was the first one specifically modelled for a HAeBe star using the MA geometry. This calibration was derived to study the prototypical HAe star UX Ori (i.e. spectral type A2, M = 3 M$_{\odot}$ and R = 3 R$_{\odot}$), and was applied by Donehew \& Brittain (\cite{DonBrit11}) to HAeBes with a large spread in stellar properties. The photometric Balmer excesses of 38 HAeBe stars were modelled in paper II using again the methodology in Muzerolle et al. (\cite{Muzerolle04}), this time taking into account the specific stellar properties of each star in the sample. Despite that using photometry instead of spectroscopy reduces the accuracy of the estimates, those are the most reliable accretion rates derived for a wide sample of HAeBe stars to date, since they take into account the strong dependence of the Balmer excess-accretion rate calibration on the stellar properties. Very recently, Pogodin et al. (\cite{Pogodin12}) reported accretion rates for a few HAeBes, obtained from spectra taken with the XShooter spectrograph on the VLT. The great capabilities of this instrument could be
further exploited, given that the mentioned dependences on the stellar properties were overlooked by using the Balmer excess-accretion rate
calibration referred in Donehew \& Brittain (\cite{DonBrit11}), which is strictly valid only for a very specific set of stellar parameters (Muzerolle et al. \cite{Muzerolle04}).

\section{Accretion tracers}
\label{Sect: Accretiontraz}

\begin{figure*}
%[!hbtp]
\centering
 \includegraphics[width=160mm,clip=true]{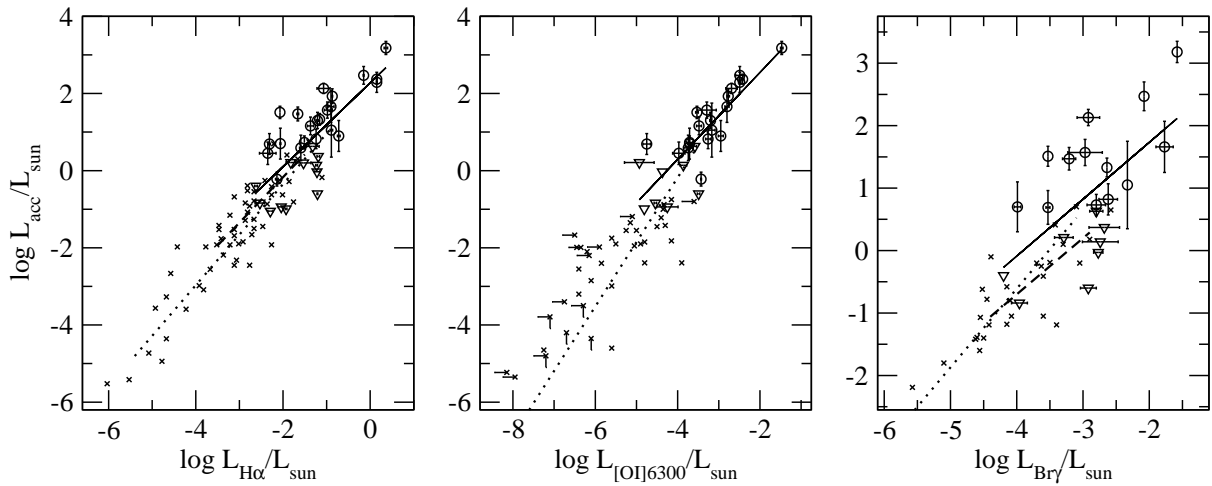}
\caption{Accretion luminosity against the H$\alpha$, [\ion{O}{i}]6300 and Br$\gamma$ luminosities . HAeBes are represented with circles and triangles for the upper limits on the accretion rates. Data for lower mass stars (crosses, upper limits are indicated with bars, when provided) were taken from Fang et al. (\cite{Fang09}), Herczeg \& Hillenbrand (\cite{Herczeg08}), and Calvet et al. (\cite{Calvet04}). Dotted and dashed lines are the empirical calibrations for low and intermediate-mass TT stars, and solid lines are the fits for the HAeBes. Figure from Paper II.}
\label{Figure:emcor}
\end{figure*}

A major result that is being settled is that several emission lines that are commonly used to estimate accretion rates for TT stars (Rigliaco et al. \cite{Rigliaco11}) are also valid to derive accretion rates for the HAeBes, under very similar empirical expressions. This has been shown and confirmed for the Br$\gamma$ line in Donehew \& Brittain (\cite{DonBrit11}) and Paper II (see also van den Ancker \cite{vandenancker05}), and has also been proven for the H$\alpha$, and [OI]6300 lines (Fig. \ref{Figure:emcor}). The origin of the correlations between the accretion and emission line luminosities is not yet clear, and several ongoing XShooter programs aim to know if the correlations can be extended to other emission lines or to early-type HAeBes. As an example of the caution that has to be taken, the width of the H$\alpha$ line at 10$\%$ of peak intensity is a valid accretion tracer for TT stars (White \& Basri, \cite{White03}, Natta et al. \cite{Natta04}), but cannot be generally applied to derive accretion rates in HAeBe stars (Paper I, Paper II). One mayor reason behind the break of the correlation between the mass accretion rate and the H$\alpha$ line width is that the high rotational velocities of HAeBe stars make the H$\alpha$ line wider, regardless of the value of $\dot{M}_{\rm acc}$, which is in turn consistent with MA (Paper II).   

\section{Accretion evolution}
\label{Sect: Accretionevol}

The timescale when gaseous planets could be formed is constrained by the gas remnant in the circumstellar disks that surround young stars. Even though H$_{2}$ is the major constituent, its direct detection is far away from being easy (Richter et al. \cite{Richter02}). Other gas probes provide estimates with various degrees of uncertainty (Kamp et al. \cite{Kamp11}). An alternative to test the inner gas dissipation timescale is by comparing the accretion rates and the stellar ages shown by PMS stars. In this way, a power-law decay provides an exponent in between 1.2 and 1.5 for T Tauri stars (Hartmann et al. \cite{Hartmann98}, Caratti o Garatti et al. \cite{Caratti12}). The same type of fit provides a steeper decline, with an exponent in between 1.1 and 3.2 for HAeBe stars (Paper III). Alternatively, an exponential decay provides a dissipation timescale of $\sim$ 2.3 Myr for TTs (Fedele et al. \cite{Fedele10}), whereas HAeBes yield $\sim$ 1.3 Myr. These results suggest that the inner gas dissipates faster around HAeBe stars, which could have implications on the physical process able to form planets on shorter timescales. It must be noted that the values mentioned were derived considering the HAeBe regime as a whole, but further analysis should be carried out taking into account smaller stellar mass bins. The dependences of both $\dot{M}_{\rm acc}$ and the stellar age on the stellar mass, the most massive HAeBes being the youngest and strongest accretors (Paper II), are most probably bringing in strong uncertainties on the values obtained (Paper III).

%\begin{figure}
%\centering
%\includegraphics[width=80mm,clip=true]{Macc_final5.eps}
%\caption{...}
%\end{figure} 
\section{Accretion, spectral energy distributions and disk masses}
\label{Sect: Accretioparam}

The dust structure of disks surrounding PMS stars can be constrained from the analysis of the spectral energy distributions (SEDs). A major achievement of the \emph{Spitzer} space telescope has been the characterization of a wide number of ''transitional'' disks, whose SEDs show an IR excess that starts at around 10 $\mu$m. This longer wavelength, compared to the corresponding one for ''classical'' SEDs, points to a depletion of the small dust grains close to the central star. Since a given IR wavelength probes a specific dust temperature, the division between transitional and classical disks in HAeBes could be defined considering a shorter wavelength, because this traces regions farther away from the central object than the same wavelength for -colder- TT stars. With this caveat in mind, it was found that HAeBes whose IR excess starts at around 2.2 $\mu$m or longer wavelengths (group $K$ sources) tend to show mass accretion rates $\sim$ 10 times lower than the remaining objects -with IR excesses starting at around 1.5 $\mu$m; group $JH$- (Paper III). This result is analogous to that found when transitional and classical T Tauri stars are compared (Najita, Strom \& Muzerolle \cite{Najita07}), pointing that both gas and dust dissipate at a similar rate for most PMS objects. In addition, the intrinsic near IR colour excess, that samples the dust in the inner disk, correlates with $\dot{M}_{\rm acc}$ for ''active'' T Tauris ($\dot{M}_{\rm acc}$ $\geq$ 10$^{-8}$ M$_{\odot}$ yr$^{-1}$) and HAeBes ($\dot{M}_{\rm acc}$ $\geq$ 10$^{-7}$ M$_{\odot}$ yr$^{-1}$). This is in agreement with viscous disk models (Meyer, Calvet, \& Hillenbrand \cite{Meyer97}, Calvet et al. \cite{Calvet00}, van den Ancker \cite{vandenancker05}, Paper III).\\

\begin{figure}
\centering
\includegraphics[width=80mm,clip=true]{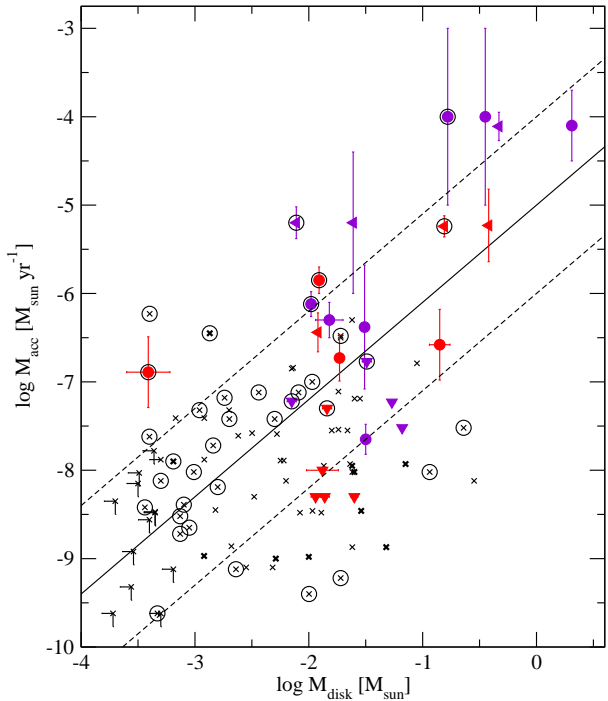}
\caption{Mass accretion rate against disk mass. Crosses are T Tauri stars from Hartmann et al. (\cite{Hartmann98}) and Najita et al. (\cite{Najita07}). Dark crosses are transitional disks. The HAeBes are represented with solid symbols, triangles for upper limits. Group $JH$ stars are in violet and group $K$ stars in red. The solid line represents the best fit to the HAeBes (M$_{\rm disk}$ $\propto$ $\dot{M}_{\rm acc}$$^{1.1 \pm 0.3}$) and the dashed lines $\pm$ 1 dex. The presence of close binaries is indicated with circles surrounding the symbols. Figure from Paper III.}
\label{Fig:Mdisk_Macc}
\end{figure}

Circumstellar dust and gas dissipation can be caused by different physical processes, such as grain growth, photoevaporation or planet formation, that open gaps in the inner parts of the disks (Cieza \cite{Cieza08}). The physical process that dominates disk disspation could be constrained from the relative position of evolved disks with respect classical ones in the $\dot{M}_{\rm acc}$ versus M$_{\rm disk}$ plane (Alexander \& Armitage \cite{Alexander07}). Following the work in Najita et al. (\cite{Najita07}) for TT stars, in Paper III the accretion and disk mass properties of HAeBes were compared (Fig. \ref{Fig:Mdisk_Macc}). Disk masses derived from mm emission scale with the mass accretion rate roughly as M$_{\rm disk}$ $\propto$ $\dot{M}_{\rm acc}$$^{1}$ also for the intermediate-mass regime, which is again the expected from viscous dissipation models (Hartmann et al. \cite{Hartmann98}, Dullemond, Natta \& Testi \cite{Dullemond06}, Najita et al. \cite{Najita07}). However, disk masses are typically smaller in group $K$ than in group $JH$, which contrasts with the larger disk masses found for transitional T Tauri stars, when compared to classical ones. This could be explained from a different physical mechanism dominating disk dissipation, depending on the stellar mass. In particular, the low accretion rates and large disk masses typically shown by transitional TT stars would be consistent with ongoing formation of giant planets in the inner disks, which dissipates the gas content, and at the same time need from a large mass reservoir from the disk to be formed. On the other hand, the low accretion rates and low disk masses shown by most group $K$ HAeBe stars could probably be better explained from photoevaporation that sweeps out both the gas and dust content in their disks. This difference between TT and HAeBe stars should be further tested using larger samples, and considering the strong uncertainties involved in the calculation of disk masses (Andrews \& Williams \cite{AndrewsWilliams07}, Paper III).

\section{Conclusions}
\label{Sect: Conclusions}

Magnetospheric accretion modelling also reproduces the observed excess in the Balmer region of the spectra of HAe and late-type HBe stars. The accretion rate estimates derived in this way correlate with the luminosity of several emission lines that are used as accretion tracers for low mass T Tauri star, following similar empirical expressions. These are crucial to carry out future accretion studies considering a large number of PMS stars.\\

The comparison between the accretion rates and several stellar and disk parameters provide similar trends for both T Tauri and HAeBe stars, suggesting common physical processes -viscous accretion- in order to explain them. However, there are two major differences between TTs and HAeBes. First, the inner gas dissipation timescale, derived by relating the accretion rates and the stellar ages, seems to be slightly faster for massive stars. Second, the relative position in the $\dot{M}_{\rm acc}$-M$_{\rm disk}$ plane suggests that a different physical process dominates disk dissipation depending on the stellar mass. That plane is a potential tool to analyze the incidence of planet formation as a dominating disk dissipation mechanism, whose results could be linked to those coming from the statistics on already formet planets around MS stars. Moreover, since these studies are mainly restricted to low-mass, slowly rotating MS objects (Lagrange et al. \cite{Lagrange09}), the $\dot{M}_{\rm acc}$-M$_{\rm disk}$ plane reveals as a unique tool to statistically study the incidence of planet formation around the most massive PMS stars eventually able to form planets in their circumstellar disks, the HAeBes.

\acknowledgements
This work was partially supported by Spanish grant AYA-2008 01727.

% Use this code if you wish to generate your bibliography with BibTeX;
% please replace first the string "an-demo" below with the name(s) of
% the BibTeX data base(s) you want to use.
% The resulting bibliography-output (the contents of the .bbl file)
% must be pasted into this file before submission.
%
% \bibliographystyle{an}
% \bibliography{an-demo}
%
% Replace the following example bibliography with your references
% before submission:

\end{document}